\documentclass{aastex}
\usepackage{spr-astr-addons}
\usepackage{graphicx}
\usepackage{txfonts}
\usepackage{xcolor}
\usepackage{tikz}
\usepackage{rotating}
\usepackage{lscape}
\usepackage{adjustbox}
\usepackage[version=4]{mhchem}
\usepackage[normalem]{ulem}
\usepackage{float}
\usepackage{natbib}
\pdfoutput=1 
\usepackage{amsmath,amstext,amssymb}
\usepackage[T1]{fontenc}
\usepackage[colorlinks=true, allcolors=blue]{hyperref}

\RequirePackage{color}
\begin{document}
	
	\title{Identification of interstellar cyanamide towards the hot molecular core G358.93--0.03 MM1}
	\shorttitle{Cyanamide towards G358.93--0.03 MM1}
	\shortauthors{Manna \& Pal}

	\author{Arijit Manna\altaffilmark{1}} \and \author{Sabyasachi Pal\altaffilmark{1}}
	\email{arijitmanna@mcconline.org.in}
	
	\altaffiltext{1}{Department of Physics and Astronomy, Midnapore City College, Paschim Medinipur, West Bengal, India 721129 \\email: {arijitmanna@mcconline.org.in}}
	
	\begin{abstract}
The amide-related molecules are essential for the formation of the other complex bio-molecules and an understanding of the prebiotic chemistry in the interstellar medium (ISM). We presented the first detection of the rotational emission lines of the amide-like molecule cyanamide (\ce{NH2CN}) towards the hot molecular core G358.93--0.03 MM1 using the Atacama Large Millimeter/Submillimeter Array (ALMA). Using the rotational diagram model, the derived column density of \ce{NH2CN} towards the G358.93--0.03 MM1 was (5.9$\pm$2.5)$\times$10$^{14}$ cm$^{-2}$ with a rotational temperature of 100.6$\pm$30.4 K. The derived fractional abundance of \ce{NH2CN} towards the G358.93--0.03 MM1 with respect to \ce{H2} was (4.72$\pm$2.0)$\times$10$^{-10}$, which is very similar to the existent three-phase warm-up chemical model abundances of \ce{NH2CN}. We compare the estimated abundance of \ce{NH2CN} towards G358.93--0.03 MM1 with other sources, and we observe the abundance of \ce{NH2CN} towards G358.93--0.03 MM1 is nearly similar to that of the sculptor galaxy NGC 253 and the low-mass protostars IRAS 16293--2422 B and NGC 1333 IRAS4A2. We also discussed the possible formation mechanisms of \ce{NH2CN} towards the hot molecular cores and hot corinos, and we find that the \ce{NH2CN} molecule was created in the grain-surfaces of G358.93--0.03 MM1 via the neutral-neutral reaction between \ce{NH2} and CN.
		
	\end{abstract}
	
	
	\begin{table*}
		\centering
		\caption{Observation summary of G358.93--0.03 using the ALMA band 6. }
		\begin{adjustbox}{width=1.0\textwidth}
			\begin{tabular}{ccccccccccccccccc}
				\hline 
				Band&Date of observation& Start time&End time &Number  &Frequency range&Spectral resolution & Field of View (FOV)\\
				&(YYYY MMM DD)       &(hh:mm) &  (hh:mm)      &  of antennas                  & (GHz)	    & (kHz)              &($\prime\prime$)\\
				\hline
				6&	2019 Oct 07        &22:01&22:23&42&216.44--217.38&488.28&26.599\\
				&                     &     &     &   &218.08--218.55&282.23&    \\  
				&                     &     &     &   &219.88--220.35&282.23&    \\
				&                     &     &     &   &220.33--220.80&282.23&    \\
				
				\hline
			\end{tabular}
		\end{adjustbox}	
		\label{tab:obs}
	\end{table*}
	
	\section{Introduction} 
The amide-type molecule cyanamide (\ce{NH2CN}) was formed with amide and cyanide functional groups, which acted as a possible precursor of urea (\ce{NH2CONH2}) in the ISM. The urea was formed via the hydrolysis of \ce{NH2CN} in the ISM \citep{kil47}. The carbodiimide (HNCNH) molecule is known as one of the isomer of \ce{NH2CN}, which contains the --NCN-- frame \citep{tur75, cou18}. The molecules with the --NCN-- frame play a major role in linking biological processes and in the assembly of amino acids into peptides \citep{wil81}. The HNCNH molecule forms due to photochemical and thermally induced reactions of \ce{NH2CN} in interstellar ice analogues \citep{duv05}. Except for urea, the \ce{NH2CN} molecule also acts as a possible precursor of 2-amino-oxazole, cytosine, 3'-cyclic phosphate (pyrimidine ribonucleotide), and beta-ribocytidine-2 via biochemistry synthesis routes towards the hot cores/corinos \citep{jim20}. The \ce{NH2CN} molecule contains 0$^{+}$ and 0$^{-}$ substates with a pyramidal equilibrium structure \citep{sharma21}. The dipole moments of the 0$^{+}$ and 0$^{-}$ substrates are $\mu_{a}=4.25\pm0.02$ D and $\mu_{a}=4.24\pm0.02$ D, respectively \citep{bro85, sharma21}. In the ISM, evidence of \ce{NH2CN} was found towards the hot molecular cores Sgr B2 (N), Orion KL, IRAS 20126+410, and NGC 6334I \citep{tur75, whi03, pa17, lig20}. The rotational emission lines of \ce{NH2CN} were also detected towards the low-mass solar-like protostars IRAS 16293--2422 B and NGC 1333 IRAS2A using the ALMA and Plateau de Bure Interferometer (PdBI) \citep{cou18}. Outside of the Milky Way, evidence of the \ce{NH2CN} molecule was also found towards the sculptor galaxy NGC 253 and the starburst galaxy M82 \citep{mar06, ala11}. Recently, the emission lines of \ce{NH2CN} were also detected towards the hot molecular core G10.47+0.03 using the ALMA band 4, with an estimated column density of 6.60$\times$10$^{15}$ cm$^{-2}$ and a rotational temperature of 201.2 K \citep{man22a}.
	
The search for complex organic molecules using the millimeter and submillimeter telescopes in the star-formation regions gives us an idea about the chemistry in the ISM. The hot molecular cores are one of the highly chemically rich phases that are located in the high-mass star-formation regions, and they play an important role in increasing the chemical complexity in the ISM \citep{shi21}. The hot molecular cores are the ideal astronomical objects in the ISM to study the different complex and prebiotic molecular lines because they have a warm temperature ($\geq$100 K), a high gas density ($n_{\ce{H2}}$$\geq$10$^{6}$ cm$^{-3}$), and a small source size ($\leq$0.1 pc) \citep{van98}. The chemistry of the hot core is characterised by the sublimation of ice mantles, which accumulate during the star-formation activities \citep{shi21}. The gaseous molecules and atoms are frozen onto the dust grains of cold molecular clouds and prestellar cores. When the temperature increases due to star-formation activities, the chemical interaction between heavy compounds becomes active on the grain surfaces, and the complex organic molecules are formed in the star-formation regions \citep{gar06, gar13, shi21}. So, the highly dense, warm, and chemically rich abundant molecules around the protostars produce the strongest molecular line emitters, known as the hot molecular cores \citep{shi21}. The period of the hot molecular cores is thought to last about 10$^{5}$ years (medium warm-up phase) to 10$^{6}$ years (slow warm-up phase) \citep{van98,gar06, gar13}.

	\begin{figure*}
		\centering
		\includegraphics[width=1.0\textwidth]{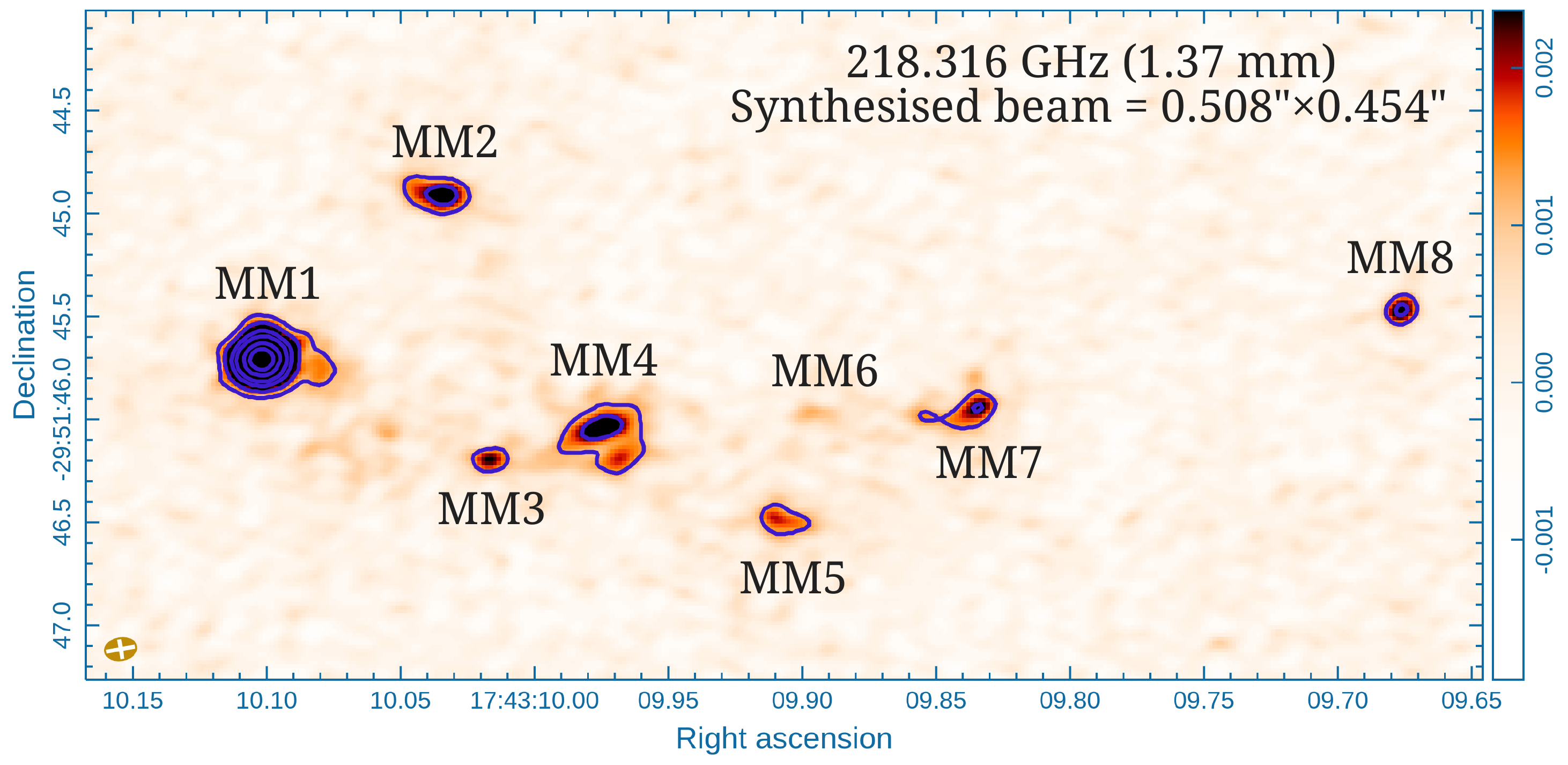}
		\caption{Sub-millimeter wavelength (1.37 mm) continuum emission image of high-mass star-formation region G358.93--0.03. The angular resolution of the continuum image was 0.508$^{\prime\prime}$$\times$0.454$^{\prime\prime}$. In the image, MM1 to MM8 are the sub-millimeter continuum sources situated in the massive star-formation region G358.93--0.03. The contour (blue colour) levels started at 3$\sigma$ where $\sigma$ = 16.40 $\mu$Jy is the RMS of the continuum emission image, and the contour level increased by a factor of $\surd$2. The bandwidth of the image was 0.47 GHz.}
		\label{fig:cont}
	\end{figure*}

The hot molecular core G358.93--0.03 MM1 is located in the high-mass star-formation region G358.93--0.03. The high-mass star-formation region G358.93--0.03 is located at a distance of 6.75$\,{}\,^{\,+0.37}_{\,-0.68}$ kpc from the earth \citep{re14,bro19}. The luminosity of the high-mass star-formation region G358.93--0.03 was $\sim$7.660$\times$10$^{3}$\textup{L}$_{\odot}$ with total gas mass 167$\pm$12\textup{M}$_{\odot}$ \citep{bro19}. The high-mass star-formation region G358.93--0.03 has eight sub-millimeter wavelength continuum sources, i.e., G358.93--0.03 MM1 to G358.93--0.03 MM8. Among the eight sources, G358.93--0.03 MM1 and G358.93--0.03 MM3 emit the molecular line emission \citep{bro19}. That means G358.93--0.03 MM1 and G358.93--0.03 MM3 are hot molecular cores because those two sources emit both continuum and line emission \citep{bro19, bay22}. Earlier, \cite{bro19} claimed that G358.93--0.03 MM1 is more chemically rich than G358.93--0.03 MM3. Earlier, the maser lines of methanol (\ce{CH3OH}), durated water (HDO), and isocyanic acid (HNCO) were detected towards the G358.93--0.03 MM1 using the ALMA, TMRT, and VLA radio telescopes \citep{bro19, chen20}. The emission lines of methyl cyanide (\ce{CH3CN}) were also detected towards the G358.93--0.03 MM1 and the G358.93--0.03 MM3 using the ALMA \citep{bro19}.

In this article, we present the first detection of the rotational emission lines of \ce{NH2CN} towards the hot molecular core region G358.93--0.03 MM1 using the ALMA band 6. The ALMA observations and data reduction of the high-mass star-formation region G358.93--0.03 were presented in Section~\ref{obs}. The result of the detection of emission lines of \ce{NH2CN} towards the G358.93--0.03 MM1 was shown in Section~\ref{res}. The discussion and conclusion of the detection of \ce{NH2CN} towards the G358.93--0.03 MM1 were shown in Section~\ref{dis} and \ref{con}.
	
\section{Observations and data reduction}
	\label{obs}
The high-mass star-formation region G358.93--0.03 was observed using the Atacama Large Millimeter/Submillimeter Array (ALMA) band 6 (frequency range of 216.44--220.80 GHz) receiver to study the massive protostellar accretion outburst (ID: 2019.1.00768.S., PI: Crystal Brogan). The observations were conducted on October 7th, 2019 with on-source integration times of 604.80 sec. The phase centre of G358.93--0.03 was ($\alpha,\delta$)$_{\rm J2000}$ = 17:43:10.000, --29:51:46.000. The observation was carried out using the forty-two antennas, with a minimum baseline of 15 m and a maximum baseline of 783 m. During the observation, J1744--3116 was used as a phase calibrator, and J1924--2914 was used as a flux calibrator and bandpass calibrator. The observation details are shown in Table~\ref{tab:obs}.
	
We used the Common Astronomy Software Application ({\tt CASA 5.4.1}) for the reduction of data using the ALMA data analysis pipeline \citep{mc07}. For flux calibration, we applied the CASA task {\tt SETJY} using the Perley-Butler 2017 flux calibrator model \citep{pal17}. For bandpass calibration and flagging the band antenna data, we applied the CASA pipeline tasks {\tt hifa\_bandpassflag} and {\tt hifa\_flagdata}. After the flux calibration, bandpass calibration, and flagging of the bad channels, we employed the task {\tt MSTRANSFORM} to separate the target object G358.93--0.03 with all available rest frequencies. We created four continuum emission images of G358.93--0.03 using the CASA task {\tt TCLEAN} with the {\tt HOGBOM} deconvolver using the frequency ranges of 216.44--217.38 GHz, 218.08--218.55 GHz, 219.88--220.35 GHz, and 220.33--220.80 GHz, respectively. Among the four individual continuum emission images, we have shown the sub-millimeter wavelength continuum emission map of the high-mass star-formation region G358.93--0.03 at frequency 218.316 GHz (1.37 mm) in Figure~\ref{fig:cont} as an example. The continuum emission image of G358.93--0.03 was created using the Cube Analysis and Rendering Tool for Astronomy (CARTA) software package \citep{cm21}. The bandwidth of this continuum emission image of G358.93--0.03 at frequency 218.316 GHz is 0.47 GHz. The effect of bandwidth smearing in those observed frequency ranges is minimal as according to \cite{mur20} the maximum bandwidth to avoid amplitude smearing between the frequency ranges of 216.44 GHz and 220.80 GHz is $\sim$1.2 GHz. From the continuum emission map of G358.93--0.03 at frequency 218.316 GHz, we identified the eight sub-millimeter wavelength continuum sources, G358.93--0.03 MM1 to G358.93--0.03 MM8. Among the eight sources, G358.93--0.03 MM1 and G358.93--0.03 MM3 are hot molecular cores because both sources emit the line emission as well as the emission lines of \ce{CH3CN} \citep{bro19}. Earlier, \citet{bro19} showed the first sub-millimeter continuum emission map of G358.93--0.03, but the authors were unable to clearly identify the continuum sources G358.93--0.03 MM6 and G358.93--0.03 MM8 due to the lower image resolution. We first individually detected all eight sub-millimeter wavelength continuum sources in the massive star-formation region G358.93--0.03 using the ALMA at wavelength 1.37 mm with an angular resolution of 0.508$^{\prime\prime}$$\times$0.454$^{\prime\prime}$. After the continuum analysis, we employed the task {\tt UVCONTSUB} to subtract the background continuum emission from the UV plane of the calibrated data. Now we created the spectral images at frequency ranges of 216.44--217.38 GHz, 218.08--218.55 GHz, 219.88--220.35 GHz, and 220.33--220.80 GHz using the task {\tt TCLEAN} with the {\tt SPECMODE} = {\tt CUBE} parameter. Finally, we used the CASA task {\tt IMPBCOR} for the correction of the synthesised beam pattern in the continuum and spectral images of G358.93--0.03.

 	\begin{table*}
	\centering
	\caption{Summary of the continuum image of G358.93--0.03 MM1.}
	\begin{tabular}{ccccccccccccccccc}
		\hline 
		Frequency & Wavelength &Integrated Flux & Peak flux &Beam size& RMS & Position angle\\
		(GHz)	 &(mm)        & (mJy)          &  (mJy beam$^{-1}$) &($^{\prime\prime}$$\times$$^{\prime\prime}$)&($\mu$Jy)&($^{\circ}$)\\
		\hline
		216.90&1.38&61.72$\pm$3.7&50.60$\pm$1.9&0.507$\times$0.456&22.12&88.879\\
		218.31&1.37&72.31$\pm$3.8&56.83$\pm$1.8&0.508$\times$0.454&16.40&87.515\\
		220.11&1.36&72.12$\pm$3.5&61.35$\pm$2.0&0.497$\times$0.455&28.23&89.169\\
		220.56&1.35&77.91$\pm$4.2&69.42$\pm$2.2&0.494$\times$0.454&22.22&86.827\\
		\hline 
	\end{tabular}	
	\label{tab:cont}
\end{table*}	
	
\section{Results}
\label{res}
\subsection{Sub-millimeter wavelength continuum emission towards the G358.93--0.03 MM1}
 	
\begin{figure*}
	\centering
	\includegraphics[width=1.0\textwidth]{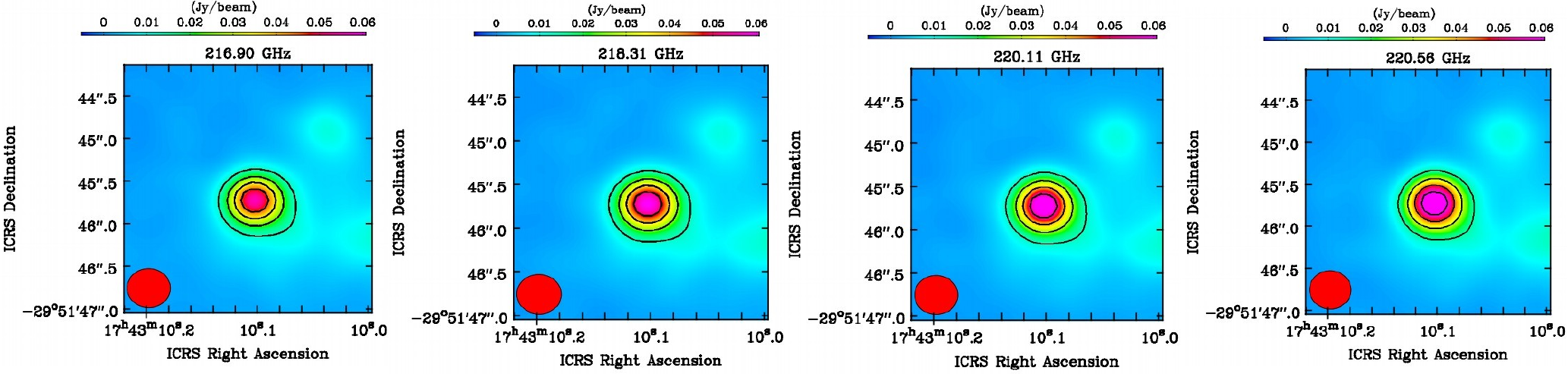}
	\caption{Sub-millimeter continuum images of G358.93--0.03 MM1 obtained with ALMA band 6 at frequencies of 216.90 GHz ($\sigma$ = 22.12 $\mu$Jy beam$^{-1}$), 218.31 GHz ($\sigma$ = 16.40 $\mu$Jy beam$^{-1}$), 220.11 GHz ($\sigma$ = 28.23 $\mu$Jy beam$^{-1}$), and 220.56 GHz ($\sigma$ = 22.22 $\mu$Jy beam$^{-1}$). The contour level started at 3$\sigma$, where $\sigma$ is the RMS of each continuum image, and the contour level increased by a factor of $\surd$2. The red circles indicate the synthesised beam of the continuum images.}
	\label{fig:gcont}
\end{figure*}

Among the eight sub-millimeter wavelength continuum sources, we mainly focus on the G358.93--0.03 MM1 ($\alpha,\delta_{\rm J2000}$ = 17:43:10.102, --29:51:45.714) because that source is the brightest sub-millimeter wavelength continuum source in the massive star-formation region G358.93--0.03, which hosts the line-rich hot core \citep{bro19, bay22}. We presented the sub-millimeter wavelength continuum emission images of the G358.93--0.03 MM1 at frequencies of 216.90 GHz (1.38 mm), 218.31 GHz (1.37 mm), 220.11 GHz (1.36 mm), and 220.56 GHz (1.35 mm) in Figure~\ref{fig:gcont}. Now we used the CASA task {\tt IMFIT} to fit the 2D Gaussian over all the continuum images and obtained an integrated flux density, peak flux density, synthesised beam size, position angle, and RMS, which are presented in Table~\ref{tab:cont}. We notice that the hot molecular core G358.93--0.03 MM1 was resolved in ALMA band 6 as it was found that the source was larger than the synthesised beam size. Earlier, \citet{bro19} also detected the millimeter-wavelength continuum emission from the eight individual continuum sources (G358.93--0.03 MM1 to G358.93--0.03 MM8) of G358.93--0.03 at frequencies of 195.58 GHz, 233.75 GHz, and 337.26 GHz, and the author mentioned the continuum emission maps of the eight individual continuum sources were resolved within the observable frequency ranges.
	
\subsection{Estimation of hydrogen (\ce{H2}) column density and dust optical depth ($\tau_\nu$) towards the G358.93--0.03 MM1}
 The peak flux density (S$_\nu$) of optically thin dust continuum emission can be expressed as,
	
	\begin{equation}
	S_\nu = B_\nu(T_d)\tau_\nu\Omega_{beam}
	\end{equation}
where $\tau_\nu$ presented the optical depth, $B_\nu(T_d)$ indicated the Planck function at dust temperature $T_d$ \citep{whi92}, and $\Omega_{beam} = (\pi/4 \ln 2)\times \theta_{major} \times \theta_{minor}$ is the solid angle of the synthesised beam. The expression of optical depth in terms of the mass density of dust can be written as,

	\begin{equation}
	\tau_\nu =\rho_d\kappa_\nu L
	\end{equation}
where $\kappa_{\nu}$ is the mass absorption coefficient, L is the path length, and $\rho_d$ indicates the mass density of dust. The mass density of the dust can be expressed in terms of the dust-to-gas mass ratio ($Z$).

	\begin{equation}
	\rho_d = Z\mu_H\rho_{H_2}=Z\mu_HN_{H_2}2m_H/L
	\end{equation}
where $\rho_{H_2}$ is the hydrogen mass density, $\mu_H$ presented the mean atomic mass per hydrogen, m$_H$ indicated the mass of hydrogen, and $N_{H_2}$ is the column density of hydrogen. We take the dust temperature $T_d$ to be 150 K \citep{chen20}, $\mu_H = 1.41$, and $Z = 0.01$ \citep{cox00}. The estimated peak flux density of the dust continuum of the G358.93--0.03 MM1 at different frequencies is presented in Table.~\ref{tab:cont}. From equations 1, 2, and 3, the column density of molecular hydrogen can be expressed as,
	
	\begin{equation}
	N_{H_2} = \frac{S_\nu /\Omega}{2\kappa_\nu B_\nu(T_d)Z\mu_H m_H}
	\end{equation}
During the estimation of the mass absorption coefficient ($\kappa_{\nu}$), we adopted the formula $\kappa_\nu = 0.90(\nu/230 GHz)^{\beta}\ cm^{2}\ g^{-1}$ \citep{moto19}, where ${k_{230} = 0.90 \ cm^{2}\ g^{-1}}$ indicated the emissivity of the dust grains at a gas density of $\rm{10^{6}\ cm^{-3}}$, which covered by a thin ice mantle at 230 GHz. We used the dust spectral index $\beta$ $\sim$ 1.7 \citep{bro19}. Using the mass absorption coefficient formula, we estimated the value of $\kappa_{\nu}$ to be 0.814, 0.823, 0.835, and 0.838 for frequencies of 216.90, 218.31, 220.11, and 220.56 GHz, respectively. We estimated the column density of hydrogen ($N_{\ce{H2}}$) for the four frequency regions towards the G358.93--0.03 MM1, which was presented in Table.~\ref{table:column density}. We take the average value to estimate the resultant hydrogen column density towards the G358.93--0.03 MM1. The estimated column density of hydrogen towards the G358.93--0.03 MM1 was (1.25$\pm$0.11)$\times$10$^{24}$ cm$^{-2}$. We also determine the value of optical depth ($\tau_\nu$) using the following equation,
	
	\begin{equation}
	T_{mb} = T_{d}(1-exp(-\tau_\nu))
	\end{equation}
where $T_{mb}$ is the brightness temperature and $T_{d}$ is the dust temperature of G358.93--0.03 MM1. For estimation of the brightness temperature, we used the Rayleigh-Jeans approximation, 1 Jy beam$^{-1} \equiv$ 118 K \citep{gor21}. The calculated dust optical depth of G358.93--0.03 MM1 with individual frequencies was presented in Table.~\ref{table:column density}. The average dust optical depth was 0.0475. The estimated dust optical depth indicated that the G358.93--0.03 MM1 is optically thin between the frequency ranges of 216.44--220.80 GHz.
	
	\begin{table}{}
	\caption{Column density of hydrogen and optical depth towards G358.93--0.03 MM1.\label{table:column density}}
	\begin{adjustbox}{width=0.47\textwidth}
		\begin{tabular}{ccccccccccccccccc}
			\hline
			{Wavelength}&{Hydrogen column density}&{Optical depth}\\
			{(mm)}&{(cm$^{-2}$)}&{($\tau_\nu$)}\\
			\hline
			1.38& $\rm{(1.28\pm0.10)\times10^{24}}$ & { 0.0397}\\
			1.37& $\rm{(1.19\pm0.30)\times10^{24}}$& { 0.0457}\\
			1.36& $\rm{(1.32\pm0.20)\times10^{24}}$ & { 0.0502}\\
			1.35& $\rm{(1.20\pm0.20)\times10^{24}}$ & { 0.0561}\\
			\hline
			{ Average Value} & $\rm{(1.25\pm0.11)\times10^{24}}$ & { 0.0475}\\
			\hline
		\end{tabular}
	\end{adjustbox}
\end{table}

\begin{figure*}
		\centering
		\includegraphics[width=1.0\textwidth]{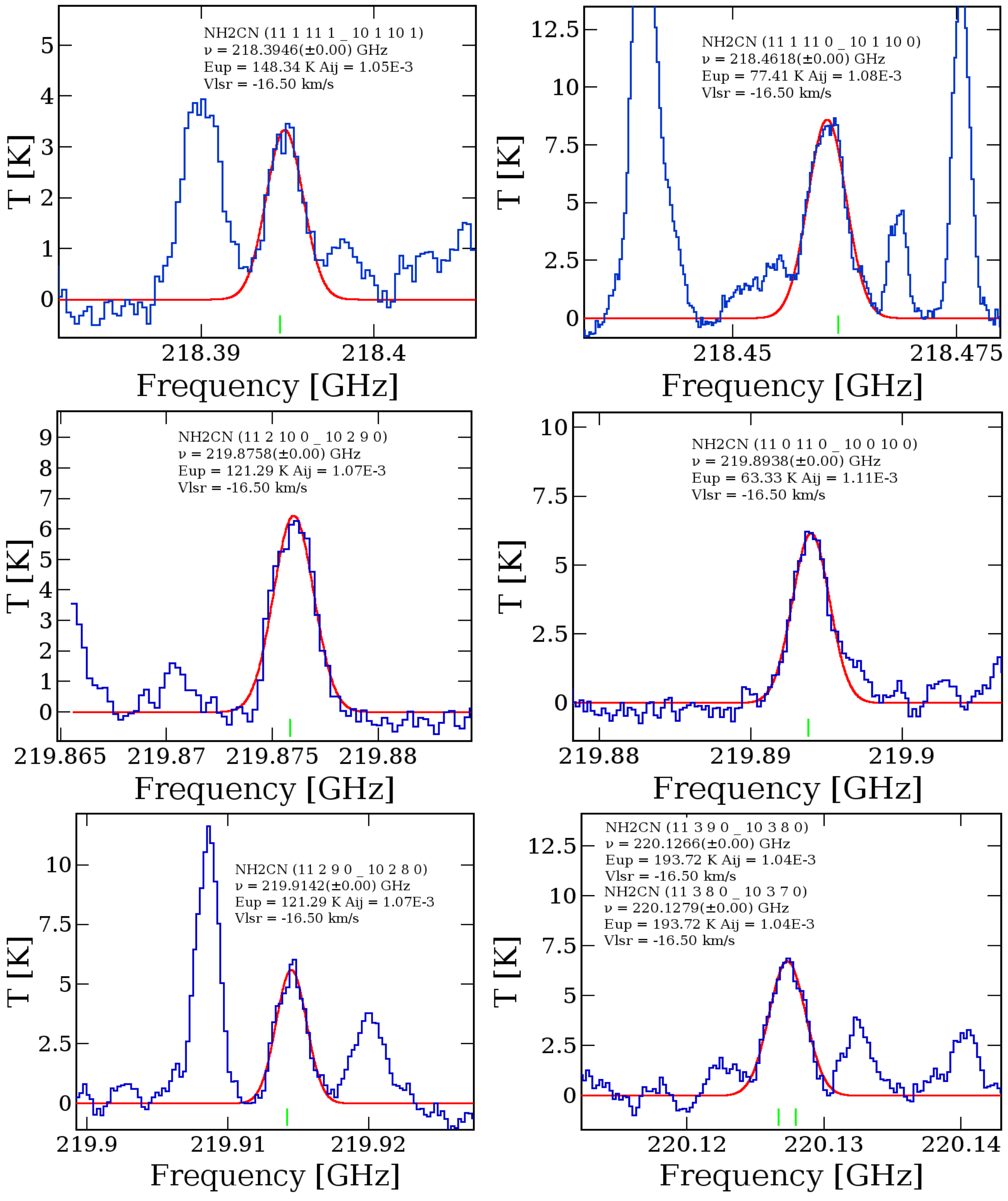}
		\caption{Rotational emission spectra of NH$_{2}$CN towards the G358.93--0.03 MM1 with different molecular transitions. The blue spectrum indicated the detected sub-millimeter spectrum of G358.93--0.03 MM1, and the red synthetic spectra indicated the Gaussian model that fitted over the detected emission lines of NH$_{2}$CN. The green vertical lines represented the rest frequency positions of the identified transitions of NH$_{2}$CN.}
		\label{fig:emission} 
\end{figure*}

	\begin{table*}
		\centering
		\caption{Summary of the Gaussian fitting line properties of the NH$_{2}$CN towards the hot molecular core G358.93--0.03 MM1.}
		\begin{adjustbox}{width=0.99\textwidth}
			\begin{tabular}{ccccccccccccccccc}
				\hline 
				Observed frequency &Transition & $E_{u}$ & $A_{ij}$ &Peak intensity&S$\mu^{2}$&FWHM & V$_{LSR}$ & $\rm{\int T_{mb}dV}$ &Remark\\
				(GHz) &(${\rm J^{'}_{K_a^{'}K_c^{'}}}$--${\rm J^{''}_{K_a^{''}K_c^{''}}}$) &(K)&(s$^{-1}$)& (K)&(Debye$^{2}$)$^{\dagger}$& (km s$^{-1}$) &(km s$^{-1}$)&(K km s$^{-1}$) & \\
				\hline
				218.394&11(1,11)--10(1,10), V = 1&148.33&1.05$\times$10$^{-3}$
				&~~3.459&199.330&3.358$\pm$0.18&--16.51$\pm$0.21&10.457$\pm$1.21&Non blended\\

				218.461&11(1,11)--10(1,10), V = 0&~~77.40&1.08$\times$10$^{-3}$&~~8.663&612.666&8.310$\pm$0.35&--16.50$\pm$0.12&47.415$\pm$0.22&Blended with \ce{NH2CHO} \\
				
				219.875&11(2,10)--10(2,9), V = 0&121.29&1.07$\times$10$^{-3}$&~~6.266&198.970&3.534$\pm$0.13&--16.49$\pm$0.94&18.059$\pm$0.22&Non blended\\
				
				219.893&11(0,11)--10(0,10), V = 0 &~~63.32&1.11$\times$10$^{-3}$&~~6.228&205.927&3.696$\pm$0.28&--16.51$\pm$0.82&18.732$\pm$0.11&Non blended\\
				
				219.914&11(2,9)--10(2,8), V = 0&121.29&1.07$\times$10$^{-3}$&~~6.025&198.996&3.728$\pm$0.25&--16.52$\pm$0.23&17.249$\pm$0.51&Non blended\\
				
				220.126&11(3,9)--10(3,8), V = 0 &193.72&1.04$\times$10$^{-3}$&~~6.861&579.630&4.508$\pm$0.17&--16.52$\pm$0.29&28.392$\pm$0.26&Non blended\\
				
				220.127&11(3,8)--10(3,7), V = 0 &193.72&10.4$\times$10$^{-3}$&~~6.861&579.624&4.506$\pm$0.36&--16.49$\pm$0.17&28.389$\pm$0.29&Non blended \\	
				\hline
			\end{tabular}
			
		\end{adjustbox}
		\label{tab:MOLECULAR DATA}
		$^{\dagger}$--The values of the S$\mu^{2}$ were taken from the JPL molecular database \citep{pic98}.
	\end{table*}

\subsection{Identification of the \ce{NH2CN} towards the G358.93--0.03 MM1}
After the production of the spectral images of G358.9--0.03 from the observable frequency ranges, we see that only the spectra of G358.93--0.03 MM1 and G358.93--0.03 MM3 show any line emission. We did not observe any line emission from other sources in G358.93--0.03. We focus to study of the complex organic molecules on the G358.93--0.03 MM1 rather than the G358.93--0.03 MM3 because the G358.93--0.03 MM1 is more chemically rich than the G358.93--0.03 MM3. We extracted the sub-millimeter wavelength molecular rich spectra from the spectral images of G358.93--0.03 to create a 0.781$^{\prime\prime}$ diameter circular region over the hot molecular core G358.93--0.03 MM1 ($\alpha,\delta_{\rm J2000}$ = 17:43:10.102, --29:51:45.714). The systematic velocity of G358.93--0.03 MM1 was --16.5 km s$^{-1}$ \citep{bro19}. We used the CASSIS \citep{vas15} with the {\tt Line Analysis} module to examine the molecular emission lines from the sub-millimeter spectra of G358.93--0.03 MM1. After the spectral analysis, we identified the rotational emission lines of amide-like molecules \ce{NH2CN} from the sub-millimeter spectra G358.93--0.03 MM1 using the Jet Population Laboratory (JPL) \citep{pic98} spectroscopic database. We detected the seven rotational emission lines of \ce{NH2CN} at frequencies ranging from 216.44 to 220.80 GHz. There are no missing transition lines of \ce{NH2CN} within the observable frequency ranges as per the JPL molecular database.

After the identification, we fitted the Gaussian model over the detected rotational emission spectra of \ce{NH2CN} using the Levenberg-Marquardt algorithm, which was available in CASSIS. After fitting a Gaussian model over the detected spectra of \ce{NH2CN}, we obtained molecular transitions ({${\rm J^{'}_{K_a^{'}K_c^{'}}}$--${\rm J^{''}_{K_a^{''}K_c^{''}}}$}), upper state energy ($E_u$) in K, line intensity ($S\mu^{2}$) in Debye$^{2}$, FWHM in km s$^{-1}$, Einstein coefficients ($A_{ij}$) in s$^{-1}$, peak intensity in K, and integrated intensity ($\rm{\int T_{mb}dV}$) in K km s$^{-1}$. After the spectral analysis, we noticed J = 11(1,11)--10(1,10) V = 1, J = 11(2,10)--10(2,9) V = 0, J = 11(0,11)--10(0,10) V = 0, J = 11(2,9)--10(2,8) V = 0, J = 11(3,9)--10(3,8) V = 0, and J = 11(3,8)--10(3,7) V = 0 transition lines of \ce{NH2CN} are not contaminated with other nearby molecular transitions. We also observed that the J = 11(1,11)--10(1,10), V = 0 transition line of \ce{NH2CN} is blended with the \ce{NH2CHO} molecular transition. During the spectral analysis, we notice the J = 11(3,9)--10(3,8) V = 0, and J = 11(3,8)--10(3,7) V = 0 transition lines of \ce{NH2CN} are present in a single spectral profile, which indicates the current spectral resolution is insufficient to resolve those two transition lines of \ce{NH2CN} towards the G358.93--0.03 MM1. The Gaussian-fitting rotational emission spectra of \ce{NH2CN} were shown in Figure~\ref{fig:emission} and a summary of the Gaussian-fitting spectral line properties of \ce{NH2CN} was presented in Table~\ref{tab:MOLECULAR DATA}.

\subsection{Spatial distribution of \ce{NH2CN} towards the G358.93--0.03 MM1}
	\label{source}
\begin{figure*}
		\centering
		\includegraphics[width=1.0\textwidth]{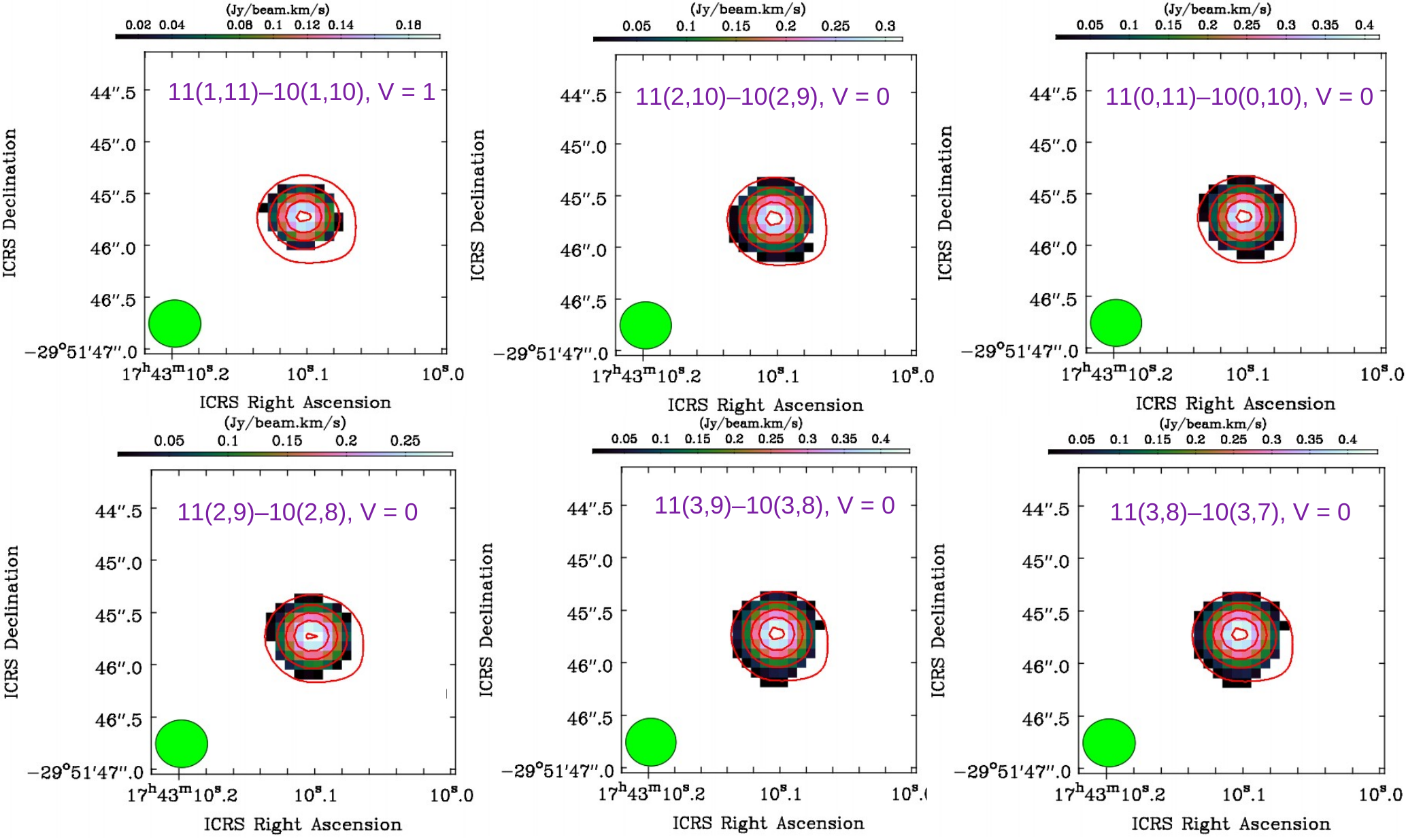}
		\caption{Integrated emission maps of NH$_{2}$CN towards the G358.93--0.03 MM1. The emission maps were overlaid with the 1.37 mm continuum emission map (red contour). The contour levels are at 20\%, 40\%, 60\%, 80\%, and 100\% of the continuum peak flux. The green circle indicated the synthesised beam of the integrated emission maps.}
		\label{fig:map}
	\end{figure*}

We created the integrated emission maps of the non-blended transitions of \ce{NH2CN} towards the G358.93--0.03 MM1 using the CASA task {\tt IMMOMENTS}. During the run of the task {\tt IMMOMENTS}, we specified the channel ranges of the spectral images where the emission lines of \ce{NH2CN} were identified. The resultant integrated emission maps of the \ce{NH2CN} towards the G358.93--0.03 MM1 were shown in Figure~\ref{fig:map}, which was overlaid on the 1.37 mm continuum emission map. We observed that the integrated \ce{NH2CN} emission maps show a peak at the position of continuum emission. The integrated emission maps indicated that the rotational emission lines of \ce{NH2CN} mainly come from the highly dense, warm inner hot core region of the G358.93--0.03 MM1. Now we used the CASA task {\tt IMFIT} for fitting the 2D Gaussian over the integrated emission maps of \ce{NH2CN} to estimate the emitting regions of \ce{NH2CN}. The deconvolved synthesised beam sizes of the \ce{NH2CN} emitting regions were derived by the following equation:\\
	
	\begin{equation}		
	\theta_{S}=\sqrt{\theta^2_{50}-\theta^2_{beam}}		
	\end{equation}
where $\theta_{beam}$ was the half-power width of the synthesised beam of the integrated emission maps and $\theta_{50} = 2\sqrt{A/\pi}$ presented the diameter of the circle whose area was surrounding the $50\%$ line peak of \ce{NH2CN} \citep{man22a,man22b}. The derived emitting regions of \ce{NH2CN} are presented in Table~\ref{tab:emittingregion}. The emitting regions of \ce{NH2CN} were found to be in the range of 0.496$^{\prime\prime}$--0.513$^{\prime\prime}$. We observed that the derived \ce{NH2CN} emitting regions are comparable to or slightly greater than the synthesised beam sizes of the integrated emission maps. That indicates the transitions of \ce{NH2CN} were not well resolved spatially or, at best, marginally resolved towards the G358.93--0.03 MM1. So it is not possible to present any conclusions about the morphology of the spatial distribution of \ce{NH2CN} towards the G358.93--0.03 MM1. High-sensitivity data with higher spatial and angular resolution observations are required to understand the spatial distribution of \ce{NH2CN} towards the hot molecular core G358.93--0.03 MM1.
	
	\begin{table}{}
	\centering
	\caption{The emitting regions of non-blended transitions of \ce{NH2CN} towards the G358.93--0.03 MM1.}
	\begin{adjustbox}{width=0.47\textwidth}
		\begin{tabular}{cccccccccccc}
			\hline
			Transition&Frequency& E$_{up}$&Emitting region\\
			
			[${\rm J^{'}_{K_a^{'}K_c^{'}}}$--${\rm J^{''}_{K_a^{''}K_c^{''}}}$]&[GHz] &[K]&[$^{\prime\prime}$]\\
			\hline
			11(1,11)--10(1,10), V = 1&218.394&148.33&0.501\\
			
			11(2,10)--10(2,9), V = 0&219.875&121.29&0.496\\
			
			11(0,11)--10(0,10), V = 0&219.893&~~63.32&0.512\\
			
			11(2,9)--10(2,8), V = 0&219.914&121.29&0.498\\
			
			11(3,9)--10(3,8), V = 0&220.126&193.72&0.513\\
			
			11(3,8)--10(3,7), V = 0&220.127&193.72&0.513\\
			\hline
		\end{tabular}	
	\end{adjustbox}

	
	\label{tab:emittingregion}
\end{table}	

	\begin{figure*}
	\centering
	\includegraphics[width=0.48\textwidth]{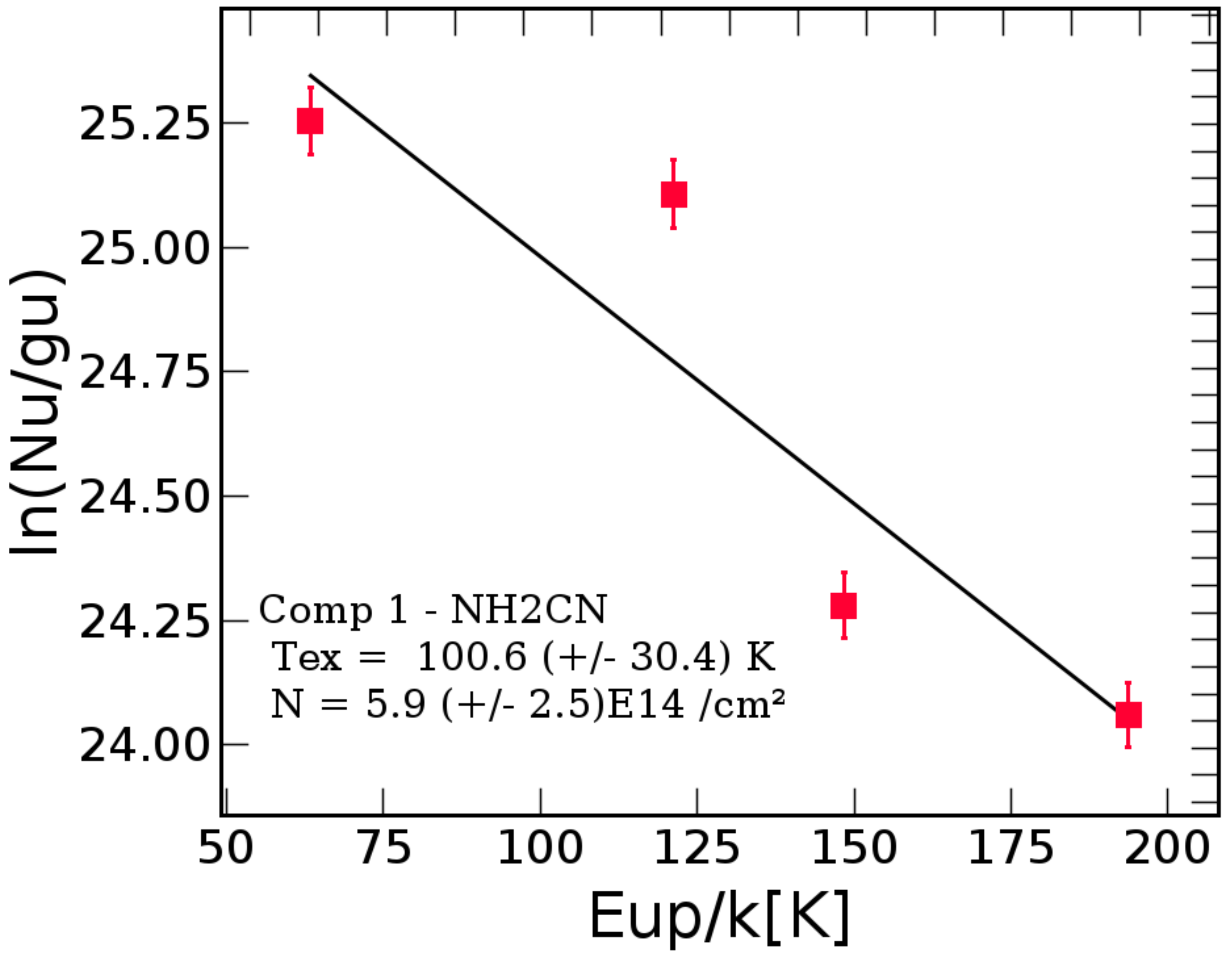}
	\caption{Rotational diagram of NH$_{2}$CN towards the G358.93--0.03 MM1. The red-filled squares indicated the values of $\ln(N_{u}/g_{u})$ which vary with different upper-state energies ($E_u$) of \ce{NH2CN}. The derived rotational temperature and column density of \ce{NH2CN} are mentioned in the figure.}
	\label{fig:rotd} 
\end{figure*}

\subsection{Rotational diagram analysis of \ce{NH2CN}}
In this work, we have identified the seven rotational emission lines of \ce{NH2CN} towards the G358.93--0.03 MM1. Rotational diagram analysis is the most efficient way to estimate the rotational temperature ($T_{rot}$) in K and the column density ($N$) in cm$^{-2}$ of the detected emission lines of \ce{NH2CN}. We used the rotational diagram model since we assumed that the observed rotational emission lines of \ce{NH2CN} are optically thin and that they are populated in local thermodynamic equilibrium (LTE) conditions. The LTE assumption was appropriate for the G358.93--0.03 MM1 because the gas density of the warm inner region of the hot core was $\sim$2$\times$10$^{7}$ cm$^{-3}$ \citep{ste21}. The equation of column density of the optically thin molecular emission lines can be expressed as \citep{gold99},
	
	\begin{equation}
	{N_u^{thin}}=\frac{3{g_u}k_B\int{T_{mb}dV}}{8\pi^{3}\nu S\mu^{2}}
	\end{equation}
where $\mu$ indicates the dipole moments of the \ce{NH2CN}, $\nu$ is the rest frequency, $S$ indicates the line strength, $g_u$ is the degeneracy of the upper state, $\int T_{mb}dV$ indicates the integrated intensity of the detected emission lines of the \ce{NH2CN}, and $k_B$ is the Boltzmann constant. The total column density of \ce{NH2CN} under the LTE condition can be expressed as,

	\begin{equation}
	\frac{N_u^{thin}}{g_u} = \frac{N_{total}}{Q(T_{rot})}\exp(-E_u/k_BT_{rot})
	\end{equation}
where, $E_u$ is the upper-state energy of \ce{NH2CN}, $T_{rot}$ is the rotational temperature of \ce{NH2CN}, and ${Q(T_{rot})}$ indicates the partition function of rotational temperature ($T_{rot}$). The partition function of \ce{NH2CN} at 75 K, 150 K, and 300 K is 1730.290, 5730.062, and 17902.30, respectively. Equation 8 can be rearranged as,

	\begin{equation}
	\ln\left(\frac{N_u^{thin}}{g_u}\right) = \ln(N)-\ln(Q(T_{rot}))-\left(\frac{E_u}{k_BT_{rot}}\right)
	\end{equation}
The equation 9 presents a linear relation between $\ln(N_{u}/g_{u})$ and upper-state energy ($E_{u}$). After the estimation of $\ln(N_{u}/g_{u})$ from equation 7, we fit a straight line to the estimated values of $\ln(N_{u}/g_{u})$, which vary with $E_{u}$. After fitting the straight line, the rotational temperature was estimated from the inverse of the slope, and the column density was estimated from the intercept of the slope. For rotational diagram analysis, the spectral line parameters of \ce{NH2CN} are estimated by fitting a Gaussian model over the observed emission lines of \ce{NH2CN}, which were presented in Table~\ref{tab:MOLECULAR DATA}. After spectral analysis, we notice only the J = 11(1,11)--10(1,10) V = 0 transition line of \ce{NH2CN} is blended with \ce{NH2CHO}, and that transition was excluded from the rotational diagram analysis. We observed that J = 11(2,10)--10(2,9) V = 0 and J = 11(2,9)--10(2,8) V = 0 transitions have the same upper-state energy of 121.29 K. During the rotational diagram, we used the J = 11(2,9)--10(2,8) V = 0 transition line because the line intensity of the J = 11(2,9)--10(2,8) V = 0 transition line is slightly higher than that of the J = 11(2,10)--10(2,9) V = 0 transition line. Similarly, we also observed that the J = 11(3,9)--10(3,8) V = 0 and J = 11(3,8)--10(3,7) V = 0 transitions of \ce{NH2CN} are present in the single spectral profile, and the upper-state energy of those two transitions is the same (193.72 K). So, we take the J = 11(3,9)--10(3,8) V = 0 transition line for the rotational diagram because the line intensity of J = 11(3,9)--10(3,8) V = 0 is slightly higher than the J = 11(3,8)--10(3,7) V = 0 transition line. The rotational diagram analysis was carried out using the J = 11(1,11)--10(1,10) V = 1, J = 11(0,11)--10(0,10) V = 0, J = 11(2,9)--10(2,8), V = 0, and J = 11(3,9)--10(3,8) V = 0 transition lines of \ce{NH2CN}. The resultant rotational diagram of \ce{NH2CN}, which was computed using the {\tt ROTATIONAL DIAGRAM} module in CASSIS, is shown in Figure~\ref{fig:rotd}. In the rotational diagram, the vertical red error bars represent the absolute uncertainty of $\ln(N_{u}/g_{u})$, which was calculated from the $\int T_{mb}dV$ of the emission spectra. From the rotational diagram, the estimated column density of \ce{NH2CN} was (5.9$\pm$2.5)$\times$10$^{14}$ cm$^{-2}$ with a rotational temperature ($T_{rot}$) of 100.6$\pm$30.4 K. The derive rotational temperature indicates that the emission lines of \ce{NH2CN} arise from the warm inner region of the G358.93--0.03 MM1 because the temperature of the hot molecular core is above 100 K \citep{van98}. The derived fractional abundance of \ce{NH2CN} with respect to \ce{H2} towards the G358.93--0.03 MM1 was (4.72$\pm$2.0)$\times$10$^{-10}$, where the \ce{H2} column density was (1.25$\pm$0.11)$\times$10$^{24}$ cm$^{-2}$.

\section{Discussion}
\label{dis}
\subsection{Comparison with other sources}
Here we compare the observed abundance of \ce{NH2CN} in G358.93--0.03 MM1 and other hot molecular cores, galaxies, high and low-mass protostars, and molecular clouds, which are shown in Table~\ref{tab:comparision}. The estimated abundance of \ce{NH2CN} towards the G358.93--0.03 MM1 is (4.72$\pm$2.0)$\times$10$^{-10}$. This means the abundance of \ce{NH2CN} towards the G358.93--0.03 MM1 is nearly similar to that of the sculptor galaxy NGC 253 and the low-mass protostars IRAS 16293--2422 B and NGC 1333 IRAS4A2. The abundance of \ce{NH2CN} towards G358.93--0.03 MM1 is approximately one order of magnitude higher as compared to the hot molecular cores Sgr B2(OH), Sgr B2(N), and Sgr B2(M), and the low-mass protostar NGC 1333 IRAS2A. Similarly, the abundance of \ce{NH2CN} towards G358.93--0.03 MM1 is approximately one order of magnitude lower as compared to the high-mass protostar IRAS 20126+4104 and the molecular cloud G+0.693. We also observed that the abundance of \ce{NH2CN} towards G358.93--0.03 MM1 is approximately two orders of magnitude lower as compared to the hot molecular core G10.47+0.03. This result indicates the formation route(s) of \ce{NH2CN} towards G358.93--0.03 MM1 is similar to NGC 253, IRAS 16293--2422 B, and NGC 1333 IRAS4A2.
	
\begin{table*}{}
	\centering
	\caption{Column density and abundance of \ce{NH2CN} towards different types of sources.}
	\begin{adjustbox}{width=1.0\textwidth}
		\begin{tabular}{cccccccccccc}
			\hline
			Source name&Type &$N$(\ce{NH2CN})&Tex&$N$(\ce{H2})&Abundance&Reference\\
			& &(cm$^{-2}$)    &(K) &(cm$^{-2}$) & & \\
			\hline
			G358.93--0.03 MM1&Hot molecular core&(5.90$\pm$2.5)$\times$10$^{14}$&100.6$\pm$30.4&(1.25$\pm$0.11)$\times$10$^{24}$&(4.72$\pm$2.0)$\times$10$^{-10}$&This work\\				
			
			G10.47+0.03&Hot molecular core&(6.60$\pm$0.1)$\times$10$^{15}$&201.2$\pm$3.3&1.30$\times$10$^{23}$&5.07$\times$10$^{-8}$&\cite{man22a}\\	
			
			Sgr B2(OH)&Hot molecular core&2.0$\times$10$^{15}$&15&5.0$\times$10$^{23}$&4.0$\times$10$^{-11}$& \cite{cu86} \\
			
			Sgr B2(N)&Hot molecular core&2.3$\times$10$^{14}$&500&3.0$\times$10$^{24}$&7.6$\times$10$^{-11}$& \cite{nu00} \\
			
			Sgr B2(M)&Hot molecular core&3.0$\times$10$^{13}$&68&2.0$\times$10$^{24}$&1.5$\times$10$^{-11}$& \cite{nu00}\\
			
			NGC 253&Sculptor Galaxy&1.2$\times$10$^{13}$&67&6.5$\times$10$^{22}$&2.0$\times$10$^{-10}$& \cite{mar06}\\
			
			IRAS 20126+4104&High-mass protostar&3.3$\times$10$^{15}$&210&2.7$\times$10$^{24}$&1.2$\times$10$^{-9}$& \cite{pa17}\\
			
			IRAS 16293--2422 B&Low-mass protostar&7.0$\times$10$^{13}$&300&1.2$\times$10$^{25}$&2.0$\times$10$^{-10}$&\cite{cou18}\\
			
			NGC 1333 IRAS2A&Low-mass protostar&2.5$\times$10$^{14}$&130&5.0$\times$10$^{24}$&5.0$\times$10$^{-11}$&\cite{cou18}\\
			
			NGC 1333 IRAS4A2&Low-mass protostar&8.5$\times$10$^{14}$&150&5.0$\times$10$^{24}$&1.7$\times$10$^{-10}$&\cite{bel20}\\	
			
			G+0.693&Giant molecular cloud&3.8$\times$10$^{13}$&6.3&1.35$\times$10$^{23}$&2.3$\times$10$^{-9}$& \cite{zh18}\\
			
			\hline
		\end{tabular}	
	\end{adjustbox}

	
	\label{tab:comparision}
\end{table*}	
	
\subsection{Possible formation and destruction pathways of \ce{NH2CN}}
In the hot core and corino objects, the formation mechanism of \ce{NH2CN} was poorly understood. According to the KIDA \citep{wak12} astrochemistry molecular reactions database, there are no gas-phase reactions that produce the abundance of \ce{NH2CN}. We found two gas-phase reactions of \ce{NH2CN} in the UMIST 2012 \citep{mce13} astrochemistry chemical network database that have the ability to synthesise \ce{NH2CN}. According to the UMIST 2012 database, the \ce{NH2CN} molecule was created via the neutral-neutral reaction between NH$_{3}$ and CN (CN + NH$_{3} \longrightarrow$ NH$_{2}$CN + H) and electron recombination of NH$_{2}$CNH$^{+}$ (NH$_{2}$CNH$^{+}$ + e$^{-} \longrightarrow$ NH$_{2}$CN + H). Earlier, \citet{cou18} also proposed another pathway for the formation of \ce{NH2CN} via the neutral-neutral reaction between \ce{NH2} and CN (NH$_{2}$ + CN $\longrightarrow$ NH$_{2}$CN). Earlier, \citet{man22a} claimed the neutral-neutral reaction between \ce{NH3} and CN is responsible for the formation of \ce{NH2CN} towards the hot molecular core G10.47+0.03. The electron recombination of NH$_{2}$CNH$^{+}$ was assumed to create \ce{NH2CN} only in 5\% of cases \citep{bel17}. Earlier, \citet{cou18} claimed the neutral-neutral reaction between \ce{NH2} and CN is the most effective pathway to produce \ce{NH2CN} on the grain surface of hot cores and corinos. Recently, \cite{zh23} also claimed the neutral-neutral reaction between \ce{NH2} and HNC is responsible for the formation of \ce{NH2CN} towards the hot corinos and hot molecular cores. Similarly, \ce{NH2CN} will be destroyed via photochemical reaction (NH$_{2}$CN + h$\nu$ $\longrightarrow$ NH$_{2}$ + CN), cosmic ray-induced photoreaction (NH$_{2}$CN + CRPHOT $\longrightarrow$ NH$_{2}$ + CN), and ion-neutral reaction (H$_{3}$$^{+}$ + NH$_{2}$CN $\longrightarrow$ NH$_{2}$CNH$^{+}$ + H$_{2}$) \citep{mce13}. Recently, \cite{zh23} also presented more destruction pathways of \ce{NH2CN} in Table 2 in the context of hot corinos and hot molecular cores.

\subsection{Previous chemical modelling of \ce{NH2CN} in the hot molecular cores}
To understand the formation mechanisms and abundance of \ce{NH2CN} towards hot molecular cores and hot corinos, \cite{cou18} used the three-phase (gas + dust + ice) warm-up chemical model using the chemical kinetics code MAGICKAL \citep{gar13}. The chemical modelling of \citet{cou18} is similar to the three-phase warm-up chemical model of \citet{gar13} which was applied towards the hot molecular cores. During the chemical modelling, \citet{cou18} used both low-density (1.6$\times$10$^{7}$ cm$^{-3}$) and high-density (6.0$\times$10$^{10}$ cm$^{-3}$) conditions. The high-density (6.0$\times$10$^{10}$ cm$^{-3}$) model is reasonable towards the hot corinos objects (IRAS 16293--2422, NGC 1333 IRAS2A, etc.) because the gas density of the hot corinos varies between $\sim$10$^{9}$ and $\sim$10$^{10}$ cm$^{-3}$ \citep{cou18}. The low-density (1.6$\times$10$^{7}$ cm$^{-3}$) model is reasonable towards the hot molecular core objects (G10.47+0.03, G31.41+0.31, Sgr B2, Orion KL, etc.) because the gas density of the hot molecular core varies between $\sim$10$^{6}$ to $\sim$10$^{8}$ cm$^{-3}$ \citep{van98, gar13}. During the chemical modelling of \ce{NH2CN} in low-density conditions, \citet{cou18} used the neutral-neutral reaction between \ce{NH2} and CN (\ce{NH2} + CN $\rightarrow$ \ce{NH2CN}) in the grain/ice chemical network to estimate the abundance of \ce{NH2CN}. During the chemical modelling of \ce{NH2CN} in low-density conditions, \citet{cou18} considered an isothermal collapse phase followed by a static warm-up phase. In the first phase (the free-fall collapse stage), the gas density rapidly increased from $\sim$1.1$\times$10$^{3}$ to 1.6$\times$10$^{7}$ cm$^{-3}$, while the dust temperature decreased from 16 K to 8 K. In the second phase (the warm-up stage), the density of gas was fixed at 1.6$\times$10$^{7}$ cm$^{-3}$ and the temperature of the dust fluctuated between 8 K and 400 K. After the modelling, \citet{cou18} estimated that the abundance of \ce{NH2CN} was 3.7$\times$10$^{-10}$ in the case of the low-density model in the warm-up stage.
		
 To ensure the chemical modelling of \cite{cou18}, we also examined another three-phase warm-up chemical modelling of \ce{NH2CN} created by \cite{zh23}. For chemical modelling of \ce{NH2CN} towards the hot molecular cores, \cite{zh23} used the NAUTILUS chemical code, which includes the gas, dust surface, and icy mantle. In the first phase (the freefall collapse stage), \cite{zh23} assumed the hot molecular cores undergoes the isothermal collapse phase at 10 K, and gas density rapidly increases from $n_{H}$ = 3$\times$10$^{3}$ cm$^{-3}$ to 1.6$\times$10$^{7}$ cm$^{-3}$. The first phase begins at $A_{V}$ = 2 mag and progresses to more than 200 mag. In the second phase (the warm-up phase), the gas density is fixed at 1.6$\times$10$^{7}$ cm$^{-3}$, and the temperature increases from 10 K to 200 K over the time scale 4$\times$10$^{5}$ yr. During the chemical modelling of \ce{NH2CN}, \cite{zh23} used the reaction \ce{NH2} + CN$\rightarrow$\ce{NH2CN} in the grain surface to estimate the abundance of \ce{NH2CN}. In the warm-up phase, \cite{zh23} observed that the abundance of \ce{NH2CN} varies between 2.0$\times$10$^{-10}$ to 6.0$\times$10$^{-9}$.

\subsection{Comparision between observed and modelled abundances of \ce{NH2CN}} 
		
To understand the formation pathways of \ce{NH2CN} towards G358.93--0.03 MM1, we compare our estimated observed abundance of \ce{NH2CN} with the modelled abundance of \ce{NH2CN} estimated by the low-density model of \cite{cou18} and \cite{zh23}. This comparison is physically reasonable because the temperature and gas density of G358.93--0.03 MM1 are $\sim$150 K \citep{chen20} and $\sim$2$\times$10$^{7}$ cm$^{-3}$ \citep{ste21}, respectively. 
Earlier, \cite{cou18} showed that the modelled abundance of \ce{NH2CN} is 3.7$\times$10$^{-10}$. Similarly, \cite{zh23} observed the abundance of \ce{NH2CN} varies between 2.0$\times$10$^{-10}$ to 6.0$\times$10$^{-9}$. We find that our estimated abundance of \ce{NH2CN} towards G358.93--0.03 MM1 is (4.72$\pm$2.0)$\times$10$^{-10}$, which is very close to the range of the modelled abundance of \ce{NH2CN} in \cite{cou18} and \cite{zh23}. This comparison indicates that \ce{NH2CN} molecule is created in the grain surfaces of G358.93--0.03 MM1 via the neutral-neutral reaction between \ce{NH2} and CN (\ce{NH2} + CN $\rightarrow$ \ce{NH2CN}).

\subsection{Searching of urea (\ce{NH2C(O)NH2}) towards the G358.93--0.03 MM1}
	
After the detection of the possible urea (\ce{NH2C(O)NH2}) precursor molecule \ce{NH2CN} from the G358.93--0.03 MM1, we also attempted to search for the rotational emission lines of \ce{NH2C(O)NH2} from the sub-millimeter spectra of the G358.93--0.03 MM1. After careful spectral analysis, we did not detect the emission lines of \ce{NH2C(O)NH2} within the limits of our LTE spectral modelling. We estimated the upper limit column density of \ce{NH2C(O)NH2} was $\leq$(1.2$\pm$0.8)$\times$10$^{13}$ cm$^{-2}$. Furthermore, we calculated the upper limit of the \ce{NH2C(O)NH2} and \ce{NH2CN} column density ratios towards the G358.93--0.03 MM1 as 0.020$\pm$0.016. For searching the emission lines of \ce{NH2C(O)NH2}, we used the Cologne Database for Molecular Spectroscopy (CDMS) spectroscopic molecular database \citep{mu05}. Our detection of the amide-like molecule \ce{NH2CN} towards the hot molecular core G358.93--0.03 MM1 using the ALMA gives more confidence that the G358.93--0.03 MM1 has the ability to create other amide-related molecules. We propose a spectral line survey of formamide (\ce{NH2CHO}), acetamide (\ce{CH3C(O)NH2}), and N-methyl formamide (\ce{CH3NHCHO}) with better spectral resolution and a higher integration time to solve the prebiotic chemistry of the amide-related molecules in the ISM.

\section{Conclusion}
\label{con}
We analysed the ALMA band 6 data of the high-mass star-formation region G358.93--0.03, and we extracted the molecular lines from the hot molecular core G358.93--0.03 MM1. The main conclusions of this work are as follows: \\\\
	1. We identified a total of seven rotational emission lines of \ce{NH2CN} towards the hot molecular core G358.93--0.03 MM1 using the ALMA band 6 observations. \\\\
	2. The estimated column density of \ce{NH2CN} towards the G358.93--0.03 MM1 was (5.9$\pm$2.5)$\times$10$^{14}$ cm$^{-2}$ with a rotational temperature of 100.6$\pm$30.4 K. The derived abundance of \ce{NH2CN} towards the G358.93--0.03 MM1 with respect to \ce{H2} was (4.72$\pm$2.0)$\times$10$^{-10}$.\\\\
	3. We compared the estimated abundance of \ce{NH2CN} towards G358.93--0.03 MM1 with other sources. We observe the abundance of \ce{NH2CN} towards G358.93--0.03 MM1 is nearly similar to that of the sculptor galaxy NGC 253 and the low-mass protostars IRAS 16293--2422 B and NGC 1333 IRAS4A2. That means the formation route(s) of \ce{NH2CN} towards G358.93--0.03 MM1 are similar to NGC 253, IRAS 16293--2422 B, and NGC 1333 IRAS4A2. \\\\
	4. We compared our estimated abundance of \ce{NH2CN} with the modelled abundances of \ce{NH2CN}, which was estimated by \citet{cou18} and \cite{zh23}. After the comparison, we found that the observed abundance of \ce{NH2CN} is nearly similar to the modelled abundances of \ce{NH2CN}. This comparison indicates that the \ce{NH2CN} molecule was created in the grain surfaces of G358.93--0.03 MM1 via the neutral-neutral reaction between \ce{NH2} and CN (\ce{NH2} + CN $\rightarrow$ \ce{NH2CN}).\\\\
	5. After the identification of the \ce{NH2CN} towards the G358.93--0.03 MM1, we also looked for the emission lines of
	urea (\ce{NH2C(O)NH2}). After the spectral analysis, we did not detect the emission lines of \ce{NH2C(O)NH2} within the limits of LTE analysis. We estimated the upper limit column density of \ce{NH2C(O)NH2} as $\leq$(1.2$\pm$0.8)$\times$10$^{13}$ cm$^{-2}$. We also derive the upper limit of the \ce{NH2C(O)NH2} and \ce{NH2CN} column density ratios towards the G358.93--0.03 MM1 as 0.020$\pm$0.016. The unsuccessful detection of \ce{NH2C(O)NH2} using the ALMA indicated that the emission lines of \ce{NH2C(O)NH2} may be below the confusion limit towards the G358.93--0.03 MM1.

\section*{Acknowledgments}{We thank the anonymous referee for the helpful comments that improved the manuscript. A.M. acknowledges the Swami Vivekananda Merit-cum-Means Scholarship (SVMCM) for financial support for this research. This paper makes use of the following ALMA data: ADS /JAO.ALMA\#2019.1.00768.S. ALMA is a partnership of ESO (representing its member states), NSF (USA), and NINS (Japan), together with NRC (Canada), MOST and ASIAA (Taiwan), and KASI (Republic of Korea), in co-operation with the Republic of Chile. The Joint ALMA Observatory is operated by ESO, AUI/NRAO, and NAOJ. }\\\\

\section*{Data availability}{The data that support the plots within this paper and other findings of this study are available from the corresponding author upon reasonable request. The raw ALMA data are publicly available at \url{https://almascience.nao.ac.jp/asax/} (project id: 2019.1.00768.S).} 
	
\section*{Funding} No funds or grants were received during the preparation of this manuscript.

\section*{Conflicts of interest}
	The authors declare no conflict of interest.
	
\section*{Author Contributions}
	S.P. conceptualize the project. A.M. analysed the ALMA data and identify the emission lines of cyanamide (\ce{NH2CN}) from G358.93--0.03 MM1. A.M. analyses the rotational diagram to derive the column density and rotational temperature of \ce{NH2CN}. A.M. and S.P. wrote the main manuscript text. All authors reviewed the manuscript.
	
	\makeatletter
	\let\clear@thebibliography@page=\relax
	\makeatother
	
\end{document}